\documentclass[12pt]{article}

\newcommand{\ie}   {{\em i.e.}}
\newcommand{\etal} {{\em et al.}}

\newcommand{\half}  {\frac{1}{2}}
\renewcommand{\bar}{\overline}
\newcommand{\VEV}[1]{\left\langle{#1}\right\rangle}
\newcommand{\ket}[1]{\vert\,{#1}\rangle}

\begin {document}
\begin{flushright}
{\small
SLAC--PUB--10233\\
October 2003\\}
\end{flushright}

\begin{center}
{{\bf\LARGE LIGHT-FRONT QUANTIZATION\\[.29ex] AND QCD
PHENOMENA}\footnote{Work supported by Department
of Energy contract DE--AC03--76SF00515.}}

\bigskip
{\it Stanley J. Brodsky \\
Stanford Linear Accelerator Center \\
Stanford University, Stanford, California 94309 \\
E-mail:  sjbth@slac.stanford.edu}
\medskip
\end{center}

\vfill

\begin{center}
{\bf\large
Abstract }
\end{center}

The light-front quantization of QCD provides an alternative to
lattice gauge theory for computing the mass spectrum, scattering
amplitudes, and other physical properties of hadrons directly in
Minkowski space. Nonperturbative light-front methods for solving
gauge theory and obtaining light-front wavefunctions, such as
discretized light-front quantization, the transverse lattice, and
light-front resolvents are reviewed.  The resulting light-front
wavefunctions give a frame-independent interpolation between
hadrons and their quark and gluon degrees of freedom, including an
exact representation of spacelike form factors, transition form
factors such as $B \to \ell \bar \nu \pi,$ and generalized parton
distributions.  In the case of hard inclusive reactions, the
effects of final-state interactions must be included in order to
interpret leading-twist diffractive contributions, nuclear
shadowing, and single-spin asymmetries. I also discuss how the
AdS/CFT correspondence between string theory and conformal gauge
theory can be used to constrain the form and power-law fall-off of
the light-front wavefunctions.  In the case of electroweak theory,
light-front quantization leads to a unitary and renormalizable
theory of massive gauge particles, automatically incorporating the
Lorentz and 't Hooft conditions as well as the Goldstone boson
equivalence theorem.  Spontaneous symmetry breaking is represented
by the appearance of zero modes of the Higgs field, leaving the
light-front vacuum equal to the perturbative vacuum.

\vfill

\begin{center}
{\it Presented at the  \\
Institute for Particle Physics Phenomenology Light Cone Workshop\\
HADRONS AND BEYOND\\
Grey College, University of Durham, Durham, England \\
 5--9 August 2003}

\end{center}

\vfill \newpage

\section{Introduction}

In Dirac's ``Front Form"\cite{Dirac:cp},  boundary conditions are
specified at a given light-front time $x^+ \equiv t + z/c; $ the
value of $x^+$  is unchanged as a light front crosses a system.
Thus, unlike ordinary time $t,$ a  moment of light-front time
$x^+=\tau$ ``stands still forever"~\cite{Montgomery}.  The
generator of light-front time translations is $P^- = i\frac{
\partial}{\partial \tau}.$  Given the Lagrangian of a quantum
field theory, $P^-$ can be constructed as an operator on the Fock
basis, the eigenstates of the free theory.  In the case of QCD,
light-front quantization provides an alternative to lattice gauge
theory for computing the mass spectrum, scattering amplitudes, and other
physical properties of hadrons directly in Minkowski space.

A remarkable advantage of light-front quantization is that the
vacuum state $\ket{0}$ of the full QCD Hamiltonian coincides with
the free vacuum.  The light-front Fock space is a Hilbert space of
non-interacting quarks and gluons, each of which satisfy $k^2 =
m^2$ and $k^- = (m^2 + k^2_\perp)/{k^+} \ge 0.$  Note that all
particles in the Hilbert space have positive energy $k^0 =
\frac{1}{2}(k^+ + k^-)$, and thus positive $k^\pm$.  Since the
plus momenta $\sum k^+_i$ is conserved by the interactions, the
perturbative vacuum can only couple to states with particles in
which all $k^+_i$ = 0; \ie, zero-mode states.   Bassetto and
collaborators~\cite{Bassetto:1999tm} have shown that the
computation of the spectrum of $QCD(1+1)$ in equal-time
quantization requires the  construction of the full spectrum of
non perturbative contributions (instantons).  In contrast, in the
light-front quantization of gauge theory (where the $k^+ = 0 $
singularity of the instantaneous interaction is defined by a
simple infrared regularization), one obtains the correct spectrum
of $QCD(1+1)$ without any need for vacuum-related contributions.

Light-front quantization can also be used to obtain a
frame-independent formulation of thermodynamics systems, such as the
light-front partition function~\cite{Brodsky:2001ww,
Alves:2002tx,Weldon:2003uz,Weldon:2003vh,Das:2003mf,Kvinikhidze:2003wc,
Beyer:2003qb}.  This application is particularly useful for
relativistic systems, such as the hadronic system produced in the
central rapidity region of high energy heavy-ion collisions.

The light-front quantization of gauge
theory~\cite{Kogut:1969xa,Tomboulis:jn,Brodsky:1997de} is usually
carried out in the light-cone gauge $A^+ = A^0 + A^z = 0$.  In
this gauge the $A^-$ field becomes a dependent degree of freedom,
and it can be eliminated from the Hamiltonian in favor of a set of
specific instantaneous light-front time interactions.  In fact in
$QCD(1+1)$ theory, the instantaneous interaction provides the
confining linear $x^-$ interaction between quarks.  In $3+1$
dimensions, the transverse field $A^\perp$ propagates massless
spin-one gluon quanta with polarization
vectors~\cite{Lepage:1980fj} which satisfy both the gauge
condition $\epsilon^+_\lambda = 0$ and the Lorentz condition
$k\cdot \epsilon= 0$.  The interaction Hamiltonian of QCD in
light-cone gauge can be derived by systematically applying the
Dirac bracket method to identify the independent
fields~\cite{Tomboulis:jn,Srivastava:2000cf}.   It contains the
usual Dirac interactions between the quarks and gluons, the
three-point and four-point gluon non-Abelian interactions, plus
instantaneous gluon exchange and quark exchange contributions.  The
renormalization constants in the non-Abelian theory have been
shown~\cite{Srivastava:2000cf} to satisfy the identity $Z_1=Z_3$
at one-loop order and are independent of the reference direction
$n^\mu$.  The QCD $\beta$ function has also been computed at one
loop~\cite{Srivastava:2000cf}.  Dimensional regularization and the
Mandelstam-Leibbrandt
prescription~\cite{Mandelstam:1982cb,Leibbrandt:1987qv,Bassetto:1984dq}
for LC gauge can be used to define the Feynman loop
integrations~\cite{Bassetto:1996ph}.  The M-L prescription has the
advantage of preserving causality and analyticity, as well as
leading to proofs of the renormalizability and unitarity of
Yang-Mills theories~\cite{Bassetto:1991ue}.  The ghosts which
appear in association with the M-L prescription from the single
poles have vanishing residue in absorptive parts, and thus do not
disturb the unitarity of the theory.  It is also possible to
quantize QCD using light-front methods in covariant Feynman
gauge~\cite{Srivastava:2000gi}.

The Heisenberg equation on the light-front is
\begin{equation}
H_{LC} \ket{\Psi} = M^2 \ket{\Psi}\ .
\end{equation}
The operator $H_{LC} = P^+ P^- - P^2_\perp,$ the ``light-cone
Hamiltonian", is frame-independent.  The Heisenberg equation can in principle be
solved by diagonalizing the matrix $\VEV{n|H_{LC}|m}$ on the free
Fock basis:~\cite{Brodsky:1997de}
\begin{equation}
\sum_m \VEV{n|H_{LC}|m}\VEV{m|\psi} = M^2 \VEV{n|\Psi}\ .
\end{equation}
The eigenvalues $\{M^2\}$ of $H_{LC}=H^{0}_{LC} + V_{LC}$ give the
squared invariant masses of the bound and continuum spectrum of
the theory.   The projections $\{\VEV{n|\Psi}\}$ of the
eigensolution on the $n$-particle Fock states are the light-front
wavefunctions.  Thus  finding the hadron eigenstates of  QCD  is
equivalent to solving a coupled many-body quantum mechanical
problem:
\begin{equation}
\left[M^2 - \sum_{i=1}^n\frac{m^2 + k^2_\perp}{x_i}\right] \psi_n
= \sum_{n'}\int \VEV{n|V_{LC}|n'} \psi_{n'}\end{equation}
where the convolution and sum is over the Fock number, transverse
momenta, plus momenta, and spin projections of the intermediate states.
The eigenvalues $M$ are the invariant masses of the complete set
of bound state and continuum solutions.

If one imposes periodic boundary conditions in $x^- = t - z/c$,
then the plus momenta become discrete: $k^+_i = \frac{2\pi}{L}
n_i, P^+ = \frac{2\pi}{L} K$, where $\sum_i n_i =
K$~\cite{Maskawa:1975ky,Pauli:1985pv}.  For a given ``harmonic
resolution" $K$, there are only a finite number of ways positive
integers $n_i$ can sum to a positive integer $K$.  Thus at a given
$K$, the dimension of the resulting light-front Fock state
representation of the bound state is rendered finite without
violating boost invariance.  The eigensolutions of a quantum field
theory, both the bound states and continuum solutions, can then be
found by numerically diagonalizing a frame-independent light-front
Hamiltonian $H_{LC}$ on a finite and discrete momentum-space Fock
basis.  Solving a quantum field theory at fixed light-front time
can thus be formulated as a relativistic extension of Heisenberg's
matrix mechanics.  The continuum limit is reached for $K \to
\infty.$ This formulation of the non-perturbative light-front
quantization problem is called ``discretized light-cone
quantization" (DLCQ)~\cite{Pauli:1985pv}.  The method preserves the
frame-independence of the Front form.

The DLCQ method has been used extensively for solving one-space
and one-time theories~\cite{Brodsky:1997de}, including
applications to supersymmetric quantum field
theories~\cite{Matsumura:1995kw} and specific tests of the
Maldacena conjecture~\cite{Hiller:2001mh}.   There has been
progress in systematically developing the computation and
renormalization methods needed to make DLCQ viable for QCD in
physical spacetime.  For example, John Hiller, Gary McCartor, and
I~\cite{Brodsky:2001ja,Brodsky:2001tp,Brodsky:2002tp} have shown
how DLCQ can be used to solve 3+1 theories despite the large
numbers of degrees of freedom needed to enumerate the Fock basis.
A key feature of our work is the introduction of Pauli Villars
fields to regulate the UV divergences and perform renormalization
while preserving the frame-independence of the theory.  A recent
application of DLCQ to a 3+1 quantum field theory with Yukawa
interactions is given in Ref.~\cite{Brodsky:2001ja} One can also
define a truncated theory by eliminating the higher Fock states in
favor of an effective
potential~\cite{Pauli:2001vi,Pauli:2001np,Frederico:2002vs}.  As
discussed below, spontaneous symmetry breaking and other
nonperturbative effects associated with the instant-time vacuum
are associated with zero mode degrees of freedom in the
light-front formalism~\cite{McCartor:hj,Yamawaki:1998cy}.

Another important nonperturbative light-front method is the
transverse
lattice~\cite{Bardeen:1979xx,Dalley:2001gj,Dalley:1998bj,Burkardt:2001mf}
which utilizes DLCQ for the $x^-$ and $x^+$ light-front
coordinates together with a spatial lattice in the two transverse
dimensions.  A finite lattice spacing $a$ can be implemented by
choosing the parameters of the effective theory in a region of
renormalization group stability to respect the required gauge,
Poincar\'e, chiral, and continuum symmetries.  For example,  Dalley
has recently computed the impact parameter dependent quark
distribution of the pion~\cite{Dalley:2003sz}.

The
Dyson-Schwinger method~\cite{Hecht:2000xa} can also be used to
predict light-front wavefunctions and hadron distribution
amplitudes by integrating over the relative $k^-$ momentum of the
Bethe-Salpeter wavefunctions to  project dynamics at $x^+ =0.$
Explicit nonperturbative light-front wavefunctions have been found
in this way for the Wick-Cutkosky model, including states with
non-zero angular momentum~\cite{Karmanov:ck}.  One can also
implement variational methods, using the structure of perturbative
solutions as a template for the numerator of the light-front
wavefunctions.  I will discuss the use of another light-front
nonperturbative method, the light-front resolvent, below.

Light-front wavefunctions are the interpolating functions between
hadrons and their quark and gluon degrees of freedom in
QCD~\cite{bro}.   For example, the eigensolution of a meson,
projected on the eigenstates $\{\ket{n} \}$ of the free
Hamiltonian $ H^{QCD}_{LC}(g = 0)$ at fixed light-front time $\tau
= t+z/c$ with the same global quantum numbers, has the expansion:
\begin{eqnarray}
\left\vert \Psi_M; P^+, {\vec P_\perp}, \lambda \right> &=&
\sum_{n \ge 2,\lambda_i} \int \Pi^{n}_{i=1} \frac{d^2k_{\perp i}
dx_i}{ 16 \pi^3}
  16 \pi^3 \delta\left(1- \sum^n_j x_j\right) \delta^{(2)}
\left(\sum^n_\ell \vec k_{\perp \ell}\right) \\[1ex]
&&\times \left\vert n; x_i P^+, x_i {\vec P_\perp} + {\vec
k_{\perp i}}, \lambda_i\right
> \psi_{n/M}(x_i,{\vec k_{\perp i}},\lambda_i)  .\nonumber
\end{eqnarray}
The set of light-front Fock state wavefunctions $\{\psi_{n/M}\}$
represents the ensemble of quark and gluon states possible when
the meson is intercepted at the light-front.  The light-front
momentum fractions $x_i = k^+_i/P^+_\pi = (k^0 + k^z_i)/(P^0+P^z)$
with $\sum^n_{i=1} x_i = 1$ and ${\vec k_{\perp i}}$ with
$\sum^n_{i=1} {\vec k_{\perp i}} = {\vec 0_\perp}$ represent the
relative momentum coordinates of the QCD constituents; the scalar
light-front wavefunctions $\psi_{n/M}(x_i,{\vec k_{\perp
i}},\lambda_i)$ are independent of the hadron's momentum $P^+ =
P^0 + P^z$, and $P_\perp$, reflecting the kinematical boost
invariance of the front form.  The physical transverse momenta are
${\vec p_{\perp i}} = x_i {\vec P_\perp} + {\vec k_{\perp i}}.$
The $\lambda_i$ label the light-front spin $S^z$ projections of
the quarks and gluons along the quantization $z$ direction.  The
physical gluon polarization vectors $\epsilon^\mu(k,\ \lambda =
\pm 1)$ are specified in light-cone gauge by the conditions $k
\cdot \epsilon = 0,\ \eta \cdot \epsilon = \epsilon^+ = 0.$ Each
light-front Fock state component then satisfies the angular
momentum sum rule: $ J^z = \sum^n_{i=1} S^z_i + \sum^{n-1}_{j=1}
l^z_j \ ; $ the summation over orbital angular momenta $ l^z_j =
-{\mathrm i} \left(k^1_j\frac{\partial}{\partial k^2_j}
-k^2_j\frac{\partial}{\partial k^1_j}\right) $ derives from the
$n-1$ relative momenta.  The numerator structure of the
light-front wavefunctions is in large part determined by the
angular momentum constraints.  Thus wavefunctions generated by
perturbation theory~\cite{Brodsky:2001ii} provides a guide to the
numerator structure of nonperturbative light-front wavefunctions.
Karmanov and Smirnov~\cite{Karmanov:ck} have formulated a covariant
version of light-front quantization by introducing a general null vector
$n^\mu, n^2=0$  to specify the light-front direction $x^+ = x \cdot n.$
All observables must be invariant under variation of $n^\mu;$  this
generalized rotational invariance provides an elegant generalization of
angular momentum on the light-front~\cite{BHHK}.

A novel way to measure the light-front wavefunction of a hadron is
to diffractively or Coulomb dissociate it into
jets~\cite{Frankfurt:it}.  Measurements by Ashery
\etal~\cite{Ashery:2003cd} of the diffractive dissociation of
pions into dijets on heavy nuclei $\pi A \to q \bar q A$ at
FermiLab show that the pion's light-front $q \bar q$ wavefunction
resembles the asymptotic solution to the evolution equation for
the pion's distribution amplitude.  The results also demonstrate
QCD color transparency -- the color dipole moment of the pion
wavefunction producing high $k_\perp$ jets interact coherently
throughout the nucleus without absorption~\cite{Bertsch:1981py}.
It would be very interesting to extend these measurements to the
diffractive dissociation of high energy protons into trijets.

Matrix elements of spacelike currents such as spacelike
electromagnetic form factors at $q^+=0$ have an exact
representation in terms of simple overlaps of the light-front
wavefunctions in momentum space with the same $x_i$ and unchanged
parton number $n$~\cite{Drell:1970km,West:1970av,Brodsky:1980zm}.
The Pauli form factor and anomalous moment are spin-flip matrix
elements of $j^+$ and thus connect states with $\Delta L_z
=1$~\cite{Brodsky:1980zm}.  Thus, these quantities are nonzero only
if there is nonzero orbital angular momentum of the quarks in the
proton.  The formulas for electroweak current matrix elements of
$j^+$ can be easily extended to the $T^{++}$ coupling of
gravitons.  In, fact, one can show that the anomalous
gravito-magnetic moment $B(0)$, analogous to $F_2(0)$ in
electromagnetic current interactions, vanishes identically for any
system, composite or elementary~\cite{Brodsky:2001ii}.  This
important feature, which follows in general from the equivalence
principle~\cite{Okun,Ji:1996kb,Ji:1997ek,Ji:1997nm,Teryaev:1999su},
is obeyed explicitly in the light-front
formalism~\cite{Brodsky:2001ii}.

The light-front Fock representation also has direct application
for the study of exclusive $B$ decays.  For example, one can write
an exact frame-independent representation of decay matrix elements
such as $B \to D \ell \bar \nu$ from the overlap of $n' = n$
parton conserving wavefunctions and the overlap of $n' = n-2$ from
the annihilation of a quark-antiquark pair in the initial
wavefunction~\cite{Brodsky:1999hn}.  The off-diagonal $n+1
\rightarrow n-1$ contributions give a new perspective for the
physics of $B$-decays.  A semileptonic decay involves not only
matrix elements where a quark changes flavor, but also a
contribution where the leptonic pair is created from the
annihilation of a $q {\bar{q'}}$ pair within the Fock states of
the initial $B$ wavefunction.  The semileptonic decay thus can
occur from the annihilation of a nonvalence quark-antiquark pair
in the initial hadron.  Intrinsic charm $\vert b \bar u c \bar c>$
states of the $B$ meson, although small in probability, can play
an important role in its weak decays because they facilitate
CKM-favored weak decays~\cite{Brodsky:2001yt}.  The ``handbag"
contribution to the leading-twist off-forward parton distributions
measured in deeply virtual Compton scattering has a similar
light-front wavefunction representation as overlap integrals of
light-front wavefunctions~\cite{Brodsky:2000xy,Diehl:2000xz}.

The distribution amplitudes $\phi(x_i,Q)$ which appear in
factorization formulae for hard exclusive processes are the
valence LF Fock wavefunctions integrated over the relative
transverse momenta up to the resolution scale
$Q$~\cite{Lepage:1980fj}.  These quantities specify how a hadron
shares its longitudinal momentum among its valence quarks; they
control virtually all exclusive processes involving a hard scale
$Q$, including form factors, Compton scattering and
photoproduction at large momentum transfer, as well as the decay
of a heavy hadron into specific final
states~\cite{Beneke:1999br,Keum:2000wi}.

The quark and gluon probability distributions $q_i(x,Q)$ and
$g(x,Q)$ of a hadron can be computed from the absolute squares of
the light-front wavefunctions, integrated over the transverse
momentum.  All helicity distributions are thus encoded in terms of
the light-front wavefunctions~\cite{Lepage:1980fj}.  The DGLAP
evolution of the structure functions can be derived from the high
$k_\perp$ properties of the light-front wavefunctions.  Similarly,
the transversity distributions and off-diagonal helicity
convolutions are defined as a density matrix of the light-front
wavefunctions.  However, it is not true that the leading-twist
structure functions $F_i(x,Q^2)$  measured in deep inelastic
lepton scattering are identical to the quark and gluon
distributions.  It is usually assumed, following the parton model,
that the $F_2$ structure function measured in neutral current deep
inelastic lepton scattering is at leading order in $1/Q^2$ simply
$F_2(x,Q^2) =\sum_q  e^2_q  x q(x,Q^2)$, where $x = x_{bj} = Q^2/2
p\cdot q$ and $q(x,Q)$ can be computed from the absolute square of
the proton's light-front wavefunction.  Hoyer,
Marchal,  Peigne, Sannino, and I have shown that this standard
identification is incomplete~\cite{Brodsky:2002ue}; one cannot
neglect the interactions which occur between the times of the
currents in the current correlator even in light-cone gauge.  For example,
the final-state interactions lead to the Bjorken-scaling diffractive
component $\gamma^* p \to p X$ of deep inelastic scattering.  Since
the gluons exchanged in the final state carry negligible $k^+$,
the Pomeron structure function closely resembles that of the
primary gluon.  The structure function of the Pomeron distribution
of a hadron is not derived from the hadron's light-front
wavefunction and thus is not a universal quantity.  The diffractive
scattering of the fast outgoing quarks on spectators in the target
in turn causes shadowing in the DIS cross section.  Thus the
depletion of the nuclear structure functions is not intrinsic to
the wave function of the nucleus, but is a coherent effect arising
from the destructive interference of diffractive channels induced
by final-state interactions.

Measurements from HERMES, SMC, and Jlab show a significant
single-spin asymmetry in semi-inclusive pion leptoproduction
$\gamma^*(q) p \to \pi X$ when the proton is polarized normal to
the photon-to-pion production plane.  Hwang, Schmidt, and
I~\cite{Brodsky:2002cx} have shown that final-state interactions
from gluon exchange between the outgoing quark and the target
spectator system lead to single-spin asymmetries in deep inelastic
lepton-proton scattering at leading twist in perturbative QCD;
{\it i.e.}, the rescattering corrections are not power-law
suppressed at large photon virtuality $Q^2$ at fixed $x_{bj}$.  The
existence of such single-spin asymmetries (the Sivers effect)
requires a phase difference between two amplitudes coupling the
proton target with $J^z_p = \pm \frac{1}{2}$ to the same
final-state, the same amplitudes which are necessary to produce a
nonzero proton anomalous magnetic moment.  The single-spin
asymmetry which arises from such final-state interactions is in
addition to the Collins effect which measures the transversity
distribution $\delta q(x,Q).$   These effects highlight the
unexpected importance of final- and initial-state interactions in
QCD observables---they lead to leading-twist single-spin
asymmetries, diffraction, and nuclear shadowing, phenomena not
included in the light-front wavefunctions of the target.
Alternatively, as discussed by Belitsky, Ji, and
Yuan~\cite{Belitsky:2002sm}, one can augment the light-front
wavefunctions by including the phases induced by initial and final
state interactions.  Such wavefunctions correspond to solving the
light-front bound state equation in an external field.

\section{Light-Front Hadron Dynamics and the AdS/CFT Correspondence}

A precise correspondence has been established between quantum
field theories and string/M-theory on Anti-de Sitter spaces
(AdS)~\cite{Maldacena:1997re}, where strings live on the curved
geometry of the AdS space and the observables of the
corresponding conformal field theory are defined on the boundary of the
AdS space.  A remarkable consequence of the AdS/CFT correspondence is the
derivation~\cite{Polchinski:2001tt} of dimensional counting rules
for the leading power-law fall-off of hard exclusive
processes~\cite{Brodsky:1973kr, Matveev:ra}.  The derivation from
supergravity/string theory does not rely on perturbation theory
and thus is more general than perturbative QCD
analyses~\cite{Lepage:1980fj}.

The corrections from nonconformal
effects in QCD are caused by quantum corrections and quark masses
and should be moderate in the ultraviolet region.
Theoretical~\cite{vonSmekal:1997is,Zwanziger:2003cf,Howe:2002rb,Howe:2003mp}
and phenomenological~\cite{Mattingly:ej,Brodsky:2002nb}  evidence
is now accumulating that the QCD coupling becomes constant at
small virtuality; \ie, $\alpha_s(Q^2)$ develops an infrared fixed
point.  Indeed, QCD appears to be a nearly-conformal theory  even at
moderate momentum transfers~\cite{Brodsky:2003rs}.

Recently Guy de T\'eramond and I have shown how counting rules for
the nominal power-law fall-off of light-front wavefunctions at
large relative transverse momentum can be obtained from the
AdS/CFT correspondence~\cite{Brodsky:2003px}.  The goal is to use
the AdS/CFT correspondence in the conformal domain to constrain
the form of the light-front wavefunctions of hadrons in QCD.  To do
this, we consider the dual string theory
at finite 't Hooft coupling
$g_s N_C,$ the product of the string  constant $g_s \sim g^2_{YM}$ and the
number of colors.  The power-law fall-off of light-front Fock-state
hadronic wavefunctions then follows from the scaling properties of string
states in the large-$r$ region of the AdS space as one approaches the
boundary from the interior of AdS space.

Consider an operator $\Psi^{(n)}_h$ which
creates an
$n$-partonic Fock state by applying n-times $a^\dagger(k^+,\vec k_{\perp
})$ to the vacuum state, creating $n$-constituent individual states with
plus momentum $k^+$ and transverse momentum $\vec k_\perp$.  Integrating
over the relative coordinates $x_i$ and $\vec k_{\perp i}$ for
each constituent,  we find the ultraviolet behavior of
$\Psi^{(n)}_h$
\begin{equation}
\Psi^{(n)}_h(Q) \sim \int^{Q^2} [d^2 \vec k_{\perp }]^{n-1} [a^\dagger(\vec
k_\perp )]^n
~\psi_{n/h}( \vec k_{\perp }) \sim Q^{-\Delta}, \label{eq:LFuv}
\end{equation}
where the  operator $a^\dagger(\vec k_{\perp })$ scales as $1/
k_{\perp }$ at large $\vec k^2_\perp.$
The string state scales as $Q^{- \Delta}$ near the AdS boundary.
The dimension of the state $\Delta$ tracks with
the number of constituents since each interpolating fermion and gauge
field operator has a minimum twist (dimension minus spin) of one.   With
the identification $\Delta = n$  (modulo anomalous dimensions), the
power-law behavior of the light-front wavefunctions for large $\vec
k_{\perp}^2$ then follows: $ \psi_{n/h}(\vec k_{\perp }) \to \left( {\vec
k_{\perp }^2} \right)^{1-n}. $

The angular momentum dependence of the
light-front wavefunctions also follow from the near-conformal
properties of the AdS/CFT correspondence~\cite{Brodsky:2003px}.  The
orbital angular momentum component of the hadron wavefunction is
constructed in terms of powers $|\ell^z_i|$ of the
$n-1$ transverse momenta
$k_{i\perp}^\pm = k_i^1 \pm i k_i^2$.  We thus can obtain a model
the hard component of the light-front wavefunction
\begin{equation}
\psi_{n/h} (x_i, \vec k_{\perp i} , \lambda_i, l_{z i}) \sim
\frac{(g_s~N_C)^{\half (n-1)}}{\sqrt {N_C}}
~\prod_{i =1}^{n - 1}
(k_{i \perp}^\pm)^{\vert l_{z i}\vert} ~
\left[\frac{ \Lambda_o}{
 {\cal M}^2 - \sum _i\frac{\vec k_{\perp i}^2 + m_i^2}{x_i} + \Lambda_o^2}
 \right] ^{n+\vert l_z \vert -1}, \label{eq:lfwfR}
\end{equation}
The scaling properties of the hadronic interpolating operator in
the extended AdS/CFT space-time theory thus determines the scaling
of light-front hadronic wavefunctions at high relative transverse
momentum.  The scaling predictions agree with the perturbative QCD
analysis given in Ref.~\cite{Ji:bw}, but  the AdS/CFT analysis is
performed at strong coupling without the use of perturbation
theory.  Remarkably, the usual perturbative
normalization factor
$(g^2_{YM} N_C)^{n-1}$ is  replaced by $(g^2_{YM}
N_C)^{{n-1}\over 2}.$  The normalization factor The near-conformal scaling properties of
light-front wavefunctions lead to a number of other predictions for QCD
which are normally discussed in the context of perturbation theory, such
as constituent counting scaling laws for the leading power
fall-off of form factors and hard exclusive scattering amplitudes
for QCD processes.  The ratio of Pauli to Dirac baryon form factor
have the nominal asymptotic form ${F_2(Q^2) / F_1(Q^2) }\sim
1/Q^2$, modulo logarithmic corrections, in agreement with the
perturbative results of Ref.~\cite{Belitsky:2002kj}.  This analysis
can also be extended to study the spin structure of scattering
amplitudes at large transverse momentum and other processes which
are dependent on the scaling and orbital angular momentum
structure of light-front wavefunctions.

\section{Spontaneous Symmetry Breaking and Light-Front Quantization}

An important question is how one implements spontaneous symmetry
breaking in the light-front, such as chiral symmetry in QCD or the
Higgs mechanism in the Standard Model.  In the case of
the Schwinger model QED(1+1), degenerate vacua arise when one
allows for a nonzero contribution $x^-$-independent contribution to
the constrained $\psi^- =
\psi^-(x^+)$ field.  Thus zero modes of auxiliary
fields distinguish the $\theta$-vacua of massless $QED(1+1)$
\cite{Yamawaki:1998cy,McCartor:2000yy,Srivastava:1999et}
corresponding to large gauge transformations.  Zero-modes are also known
to provide the light-front representation of spontaneous symmetry
breaking in scalar theories~\cite{Pinsky:1994yi}.  It is  expected that
chiral symmetry breaking in QCD  arises from a $\tau-$ independent
contribution to the constrained $\psi^-$ fields when the $u$ and
$d$ quark masses are ignored.   The existence of such vacua also
leads to new effective interactions in the light-front
Hamiltonian.

One can use light-front quantization of the $SU(2)_{W}\times
U(1)_{Y}$~\cite{gws}  standard model to obtain a new perspective
on the Higgs mechanism and spontaneous symmetry
breaking~\cite{pre4,pre5,pre6} One first separates the quantum
fluctuations from the corresponding
zero-longitudinal-momentum-mode variables and then  applies the
Dirac procedure in order to construct the Hamiltonian.  The
interaction Hamiltonian of the Standard Model can be written in a
compact form by retaining the dependent components $A^{-}$ and
$\psi_{-}$ in the formulation.  Its form closely resembles the
interaction Hamiltonian of covariant theory, except for the
presence of additional instantaneous four-point interactions.  The
resulting Dyson-Wick perturbation theory expansion based on
equal-LF-time ordering  allows one to
perform higher-order computations in a straightforward fashion.
The singularities in the noncovariant pieces of the field
propagators can be defined using the causal ML prescription for
$1/k^{+}. $  The power-counting rules in LC gauge then become
similar to those found in covariant gauge theory.  The only ghosts
which appear in the formalism are the $n \cdot k = 0$ modes of the
gauge field associated with regulating the light-cone gauge
prescription.  For example, consider the Abelian Higgs model.  The
interaction Lagrangian is
\begin{equation} {\mathcal L}= -\frac{1}{4} F_{\mu
\nu}F^{\mu \nu}+ \vert D_\mu \phi\vert^2 -V(\phi^\dagger \phi)
\end{equation}
 where
\begin{equation}
D_\mu = \partial_\mu + i e A_\mu,\end{equation}  and
\begin{equation}V(\phi)= \mu^2 \phi^\dagger \phi + \lambda(\phi^\dagger
\phi)^2,\end{equation}
 with $\mu^2 < 0, \lambda >0.$ The complex
scalar field $\phi$ is decomposed as \begin{equation}\phi(x)=
\frac{1}{\sqrt 2} v + \varphi(x) = \frac{1}{\sqrt 2}[ v + h(x) + i
\eta(x)]\end{equation}
where $v$ is the $k^+=0$ zero mode
determined by the minimum of the potential: $v^2 =
-\frac{\mu^2}{\lambda}$, $h(x)$ is the dynamical Higgs field, and
$\eta(x)$ is the Nambu-Goldstone field.  The quantization
procedure determines $\partial \cdot A = M \eta$, the 't Hooft
condition.  One can now eliminate the zero mode component of the
Higgs field $v$ which gives masses for the fundamental quantized
fields.   The $A_\perp$ field then has mass $M=e v$ and the Higgs
field acquires mass $m^2_h = 2 \lambda v^2 = - 2 \mu^2.$
Similarly, in the case of the Standard model, the zero mode of the
Higgs field couples to the  gauge boson and Fermi fields through
its Yukawa interaction.  The zero mode  can then  be eliminated
from the theory in favor of mass terms for the fundamental matter
fields in the effective theory.  The resulting masses are
identical to those of the usual Higgs implementation of
spontaneous symmetry breaking in the Standard Model.  A new aspect
of LF quantization is that the third polarization of the quantized
massive vector field $A^\mu$ with four momentum $k^\mu$ has the
form $E^{(3)}_\mu = {n_\mu M / n \cdot k}$.  Since $n^2 = 0$, this
non-transverse polarization vector has zero norm.  However, when
one includes the constrained interactions of the Goldstone
particle, the effective longitudinal polarization vector of a
produced vector particle is $E^{(3)}_{\rm eff \, \mu}= E^{(3)}_\mu
- { k_\mu \,k \cdot E^{(3)} / k^2}$ which is identical to the
usual polarization vector of a massive vector with norm
$E^{(3)}_{\rm eff }\cdot E^{(3)}_{\rm eff }= -1$.  Thus, unlike
the conventional quantization of the Standard Model, the Goldstone
particle only provides part of the physical longitudinal mode of
the electroweak particles.   The massive gauge field propagator
has well-behaved asymptotic behavior in accordance with a
renormalizable theory, and the massive would-be Goldstone fields
can be taken as physical degrees of freedom.  Spontaneous symmetry
breaking is thus implemented in a novel way when one quantizes the
Standard Model at fixed light-front time $\tau = x^+.$ The LF
vacuum remains equal to the perturbative vacuum; it is unaffected
by the occurrence of spontaneous symmetry breaking.  In effect, one
can interpret the $k^+=0$ zero mode Higgs field as an
$x^-$-independent external field, analogous to an applied constant
electric or magnetic field in atomic physics.  In this
interpretation, the zero mode is  a remnant of a Higgs field which
persists from early cosmology; the LF vacuum however remains
unchanged and unbroken.

\section{The Non-Perturbative Light-Front T-Matrix}

The light-front formalism can be used to construct the $T-$matrix
of QCD or other quantum field theories using light-front
time-ordered perturbation theory.  The application of the
light-front time evolution operator $P^-$ to an initial state
systematically generates the tree and virtual loop graphs of the
$T$-matrix in light-front time-ordered perturbation theory.  Given
the interactions of the light-front interaction Hamiltonian, any
amplitude in QCD and the electroweak theory can be computed.  At
higher orders, loop integrals only involve integrations over the
momenta of physical quanta and physical phase space $\prod
d^2k_{\perp i} d k^+_i$.  Renormalized amplitudes can be
explicitly constructed by subtracting from the divergent loops
amplitudes with nearly identical integrands corresponding to the
contribution of the relevant mass and coupling counter terms (the
``alternating denominator method")~\cite{Brodsky:1973kb}.  The
natural renormalization scheme to use for defining the coupling in
the event amplitude generator is a physical effective charge such
as the pinch scheme~\cite{Cornwall:1989gv}.  The argument of the
coupling is then unambiguous~\cite{Brodsky:1994eh}.  The DLCQ
boundary conditions can be used to discretize the phase space and
limit the number of contributing intermediate states without
violating Lorentz invariance.  This provides  an ``event amplitude
generator" for high energy physics reactions where each particle's
final state is completely labelled in momentum, helicity, and
phase.  Since one avoids dimensional regularization and
nonphysical ghost degrees of freedom, this method of generating
events at the amplitude level could provide a simple but powerful
tool for simulating events both in QCD and the Standard Model.

One can use a similar method to construct the $T$ matrix to any
given order of perturbation theory or maximal Fock number.  The
Lippmann-Schwinger method $T = H_I + H_I G T $ provides
nonperturbative resummation at any stage.  One can also use an
elementary field to project out states with specific hadronic
quantum numbers.  The zeroes of the resolvent of the projected
Green's function should determine the mass and light-front
wavefunctions of the bound states of the theory with the same
hadronic numbers as that of the elementary field.  A related method
for calculating scattering amplitudes using the  Lanczos algorithm
has been proposed by Hiller~\cite{Hiller:2000vi}.

\section{Nonperturbative Anomalous Moment Calculations }

One of the most challenging problems in quantum electrodynamics is
to compute the anomalous magnetic moment of the leptons without
recourse to perturbation theory~\cite{Feynman,Drell:1965hg}.  In
recent work, John Hiller and Gary McCartor and
I~\cite{McCartor:2003jp,BHM,Hiller:1998cv} have shown how such a
program can be implemented using light-front methods to construct
the Fock components of the physical electron nonperturbatively.
The generalized Pauli-Villars method with ghost metric fermion
fields $\psi_{PV}$ can be used to regulate the ultraviolet
divergences.  If one rewrites the theory in terms of the zero mode
fermion fields, $ \psi \pm \psi_{PV},$  then instantaneous fermion
exchange interactions do not appear in the light-front
Hamiltonian.  In addition, the constraint equation for the
zero-norm fermion field does not require inverting a covariant
 derivative.  Thus one can implement light-front
quantization of gauge theory in a covariant gauge such as the
Feynman gauge~\cite{Srivastava:2000gi}.  If one truncates the Fock
space with a maximal number of constituents $N$, the method
includes perturbative contributions to order $n=N-1.$ The
$N-$particle truncated result for the lepton anomalous moment thus has
the form
$$a_N = \sum_{i=1}^n  c_i ~ \alpha^i + \Delta \alpha^N$$ where
$\Delta$ remains dependent on the PV masses since the mass
renormalization counter term is not itself cancelled in the
$N-$particle Fock state.  This residual dependence is similar to
the factorization scale dependence which occurs when one separates hard
and soft effects in factorization analyses.  It is possible to
make nonperturbative predictions by relating  the anomalous moment
to other observables, or  to optimize the values of the cutoffs
using theoretical criteria, such as minimization of estimated
errors.

\section*{Acknowledgments}
Work supported by the Department of Energy under contract number
DE-AC03-76SF00515.  I am grateful to Simon Dalley for organizing
LC2003 and to James Stirling and his staff for their outstanding
hospitality at the Institute for Particle Physics Phenomenology in
Durham.  I also thank my collaborators, including Guy de T\'eramond, John
Hiller, Dae Sung Hwang, Volodya Karmanov, and  Gary McCartor.

\end{document}